\documentstyle[12pt,aps]{revtex}
\parskip=\medskipamount
\def\be{\begin{equation}}
\def\ee{\end{equation}}
\def\disp{\displaystyle}

\newcounter{fig}

\begin{document}

\begin{titlepage}

\begin{center}

{\LARGE \bf  Kinetics  of Anchoring of Polymer Chains on Substrates
with Chemically Active Sites

}
\vspace{0.3in}
\vspace{0.3in}

{\large \bf G.Oshanin$^1$, S. Nechaev$^{2,3}$,  A.M.Cazabat$^{4}$ and 
M.Moreau$^1$\\}

\end{center}
\vspace{0.3in}

{\sl $^1$ Laboratoire de Physique Th\'eorique des Liquides$^{*}$ (URA 765), 
Universit\'e Pierre et Marie Curie,
T.16, 4 place Jussieu, 75252 Paris Cedex 05, France
}

{\sl $^2$
Institut de Physique Nucl\'eaire,
 Division de Physique Th\'eorique$^{*}$,
91406 Orsay Cedex, France 
}

{\sl $^3$
L.D. Landau Institute for
Theoretical Physics, 117940, Moscow, Russia
}

{\sl $^{4}$ 
Laboratoire de Physique de la Mati\`ere Condens\'ee,
Coll\`ege de France, 11 place M.Berthelot, 75252 Paris Cedex 05, France\\
}

\vspace{0.3in}

\begin{abstract}

We consider dynamics of an isolated 
polymer chain with a chemically active end-bead on
 a $2D$ solid
substrate containing immobile, randomly placed 
chemically active sites (traps). 
For a particular situation
when the end-bead can be  
irreversibly trapped
 by any of these sites, 
which results in
a complete anchoring of the 
whole chain, we calculate
the time evolution of the probability $P_{ch}(t)$
that the initially non-anchored chain
remains mobile until time $t$. We find that for
relatively short chains  $P_{ch}(t)$ 
follows  at intermediate times a standard-form $2D$
 Smoluchowski-type decay law $ln P_{ch}(t) \sim - t/ln(t)$, 
which crosses over
at very large times to the fluctuation-induced 
 dependence
 $ln P_{ch}(t) \sim - t^{1/2}$, associated with
fluctuations in the spatial distribution of traps. 
We show next that for long chains
the kinetic behavior is quite different; 
here  the intermediate-time decay is of the form
$ln P_{ch}(t) \sim - t^{1/2}$, which is the
Smoluchowski-type law associated with
subdiffusive motion of the end-bead, while the long-time
fluctuation-induced decay is described by the dependence 
$ln P_{ch}(t) \sim - t^{1/4}$, stemming out of the interplay between 
fluctuations in traps distribution and 
internal relaxations of the chain.

\end{abstract}

\vspace{0.2in}


\vspace{0.2in}

\hrule \footnotesize

$^{*}$ Unit\'e de Recherche des Universit\'es Paris XI et Paris VI
associ\'ee au CNRS.

\end{titlepage}

\section{Introduction}

Understanding of polymer dynamics on solid substrates 
impacts many areas of modern
technology, including coating,
gluing, painting or lubrication. Most of 
liquids used in these material 
processing operations are either polymer 
liquids or contain polymeric 
additives. 

Meanwhile, polymer dynamics on bare substrates or
in adsorbed polymer films has been 
studied theoretically and numerically 
only for model substrates with an ideal, 
crystalline-type order \cite{ala,alb,joel}. 
However, recent experimental
studies \cite{caza,cazb,cazc,cazd,tiberg,yoon} 
of polymer monolayers spreading
on solid substrates have given an
 ample evidence that in realistic 
situations chains dynamics  is strongly influenced
 by different types of 
disorder, associated with the presence of
 contaminants, chemically active 
"hot" sites or surface roughness.  
Such a disorder is unavoidable for real 
surfaces and induces significant
 departures from the behavior predicted for 
model systems.  

In particular, studies of light 
polydimethylsiloxane (usually abbreviated as PDMS)
 molecules spreading on 
oxidized silicon wafers have demonstrated that the form of the PDMS diffusivity $D$ 
is very 
sensitive to the chemical composition of the surface, or  more specifically, 
to the presence of the silanol sites, which can 
form a hydrogen bond with 
any chain's monomer and thus temporarily anchor the chain.
Experiments reveal 
\cite{cazc} an ideal, Rouse-type behavior of the form $D \sim N^{-1}$, where  
$N$ is the number of monomeric units in a polymer, 
at low density 
of such sites.  On 
the other hand, a stronger dependence of the form $D \sim N^{-2}$ is observed 
 \cite{cazc} at higher density of the silanol sites, which behavior 
resembles of the reptative motion and stems apparently from some collective 
effects, associated with trapping of some portion of chains serving then 
as obstacles for the rest.  

Even more striking  effect of chemical 
disorder has been observed in the case of the so-called PDMS-OH polymers, i.e. 
the PDMS molecules bearing an OH-group at one or at both of the chain's 
extremities.  The OH-group can form a strong chemical bond with any of the 
silanol sites, resulting in a complete anchoring of the whole chain by one 
of its ends.  In consequence, despite the fact that the macroscopic 
spreading power of such a liquid/solid system is positive and thus favors 
complete wetting, spreading of a sufficiently thin film of the PDMS-OH 
molecules terminates at a certain moment of time due to the presence of 
chemically active trapping sites (see \cite{tiberg} and \cite{yoon}).

Surprisingly enough, dynamics of 
polymers in presence of randomly placed traps has not been addressed 
so far, 
in contrast to the well-studied theoretically problems 
of chain dynamics on the surface with randomly placed barriers or obstacles 
(see, e.g. \cite{nechaev,stauffer} and references therein) or diffusion of monomers
in a medium with traps (see for a review  \cite{weiss,kehr,4}). 
In the present paper we discuss this practically important problem 
focusing first on a simple model appropriate to the just-described 
situation with the PDMS-OH molecules deposited on a bare silicon wafer with 
silanol sites.  More specifically, we study here dynamics of a single 
polymer, modeled as an ideal Rouse chain with a chemically active end-bead (see Fig.1),
on a 
two-dimensional ideal substrate with randomly placed perfect immobile 
traps. The end-bead 
can be irreversibly trapped upon the first encounter with any of the 
traps, which results in anchoring of the whole chain. 
Dynamics of all other beads is completely unaffected by  traps.  
For this model we find explicit results for the probability $P_{ch}(t)$
that the chemically active end-bead of the chain 
does not meet any of the trap until time $t$, or, in 
other words,
that the polymer chain, which is  
unanchored at $t = 0$,  remains completely mobile until time $t$. 
Other possible situations involving, 
in particular, reversible traps or many active groups per chain, as well as the effects
of the excluded volume interactions
 will be 
discussed in the forthcoming publication.

The paper is structured as follows: In  Section 2 we describe in more detail
the model to 
be studied and introduce notations. In Section 3 we present a reminder
on trapping kinetics of a monomer particle, which allows us to 
explain some basic
ideas concerning the effect of fluctuations in traps' spatial distribution
on trapping kinetics.  
Next, in Section 4 we show how these results can be extended
to describe the anchoring kinetics of a 
Rouse polymer chain and analyze
different kinetic regimes. 
 Finally, we conclude in the Section 5 with 
a summary and discussion of our results.

\section{The model}

Consider a polymer chain deposited on a two-dimensional solid surface, 
Fig.1, and forming no loops in the direction perpendicular to the surface.
The chain consists of $N+1$ identical beads, connected into the chain by 
harmonic springs with rigidity $\chi$, $\chi = 2 T/b^2$, $T$ being the temperature of
the solid substrate and $b$ - the average distance between the beads.  
The radii  of the beads are denoted by 
two-dimensional (time-dependent)
vectors ${\bf r}_n$, $n$ being the number of the bead in 
the chain, $n = 0, ... ,N$, $N \gg 1$. We suppose that one 
of the end-beads of the chain, namely
the bead
with $n = 0$, differs from all others in 
that it contains a chemically
active group, 
while all other beads are chemically inert.
Assuming that the springs are phantom, which 
means that we discard excluded volume interactions, we have for the 
potential energy $U(\{{\bf r}_{n}\})$ of the chain
\be \label{potential}
U\Big(\Big\{{\bf r}_{n}\Big\}\Big) \; = \; \sum_{n = 0}^{N - 1} U\Big({\bf r}_{n+1}-{\bf 
r}_n\Big) \; = \;   \frac{T}{b^2}  \; \sum_{n = 0}^{N - 1} \Big({\bf 
r}_{n+1} - {\bf r}_n\Big)^2  
\ee
Effect of the excluded volume interactions on 
the anchoring kinetics, which 
can be rather important for 
two-dimensional systems, will be discussed in detail 
elsewhere. Here we will present only some brief comments
on this point at the very end of the paper.

Further on, we suppose that 
the beads experience an action of random forces, which 
 originate from chaotic, thermal vibrations of solid 
atoms around their 
lattice positions; the beads  may thus 
perform random motion along the 
surface under the constraints imposed 
by the  springs.
Simplifying the actual situation to 
some extent \cite{gomer}, we model 
these random forces as Gaussian white noise $\zeta_{n,\alpha}(t)$, 
uncorrelated in time and space, such that
\be \label{noise}
\begin{array}{l}
\overline{\zeta_{n,\alpha}(t)} \; = \; 0 \\
\vspace{0.1in}
\overline{\zeta_{n,\alpha}(t)\zeta_{n',\alpha'}(t')} \; = \; 2 \;
\eta \; T \; \delta_{n,n'} \; \delta_{\alpha,\alpha'}  \;
\delta(t-t')
\end{array}
\ee
In Eqs.(\ref{noise}) the overline stands for 
the averaging over thermal
noise, $\delta_{\alpha,\alpha'}$ and
 $\delta_{n,n'}$ are
 the Kroneker symbols,
$\alpha = x,y$ denote the Cartesian components of random 
forces, while $\eta$ is the 
friction coefficient, which is dependent on 
the height of the barrier 
against the lateral motion and the temperature 
(see \cite{gomer} for more 
details).

We suppose next that the surface (of the surface area $S$)
contains $M$ perfect, immobile traps, which 
are placed at random positions,  
which
are denoted by vectors $\{{\bf R}_{j}\}$, $j = 1, \ldots,M$.
 In what follows we will always assume
the limit $S,M \to \infty$ with the fixed mean 
density $n_{tr} = M/S$, $n_{tr} \ll 1$. Next,
we stipulate that the action of 
the traps on the chain's beads is selective: 
the traps have completely no 
effect on all the beads of the chain, (except 
for the end-bead $n = 0$), which 
means that the traps do not react with the beads 
with $n = 1, \ldots,N$ and 
do not influence their dynamics.  On contrary, the 
end-bead is trapped at 
the first encounter with any trap and gets immobilized 
anchoring the whole
chain. As we have already mentioned, from the physical 
point of view such a 
model mimics the situation with PDMS-OH molecules 
diffusing on silicon 
wafers with the silanol sites; here,  
the silanol sites, i.e. the traps,
may form strong chemical bonds with 
the OH-groups (end-bead)  
immobilizing them.  On the other hand, 
these sites form only weak hydrogen 
bonds with any other monomer of the PDMS molecule. 
These weak bonds
create an additional (small) barrier 
against the lateral motion; 
we suppose that the influence of the silanol sites
on the dynamics of the PDMS monomers can be 
accounted for by introducing
some effective  friction 
coefficient $\eta$. A non-trivial, as a matter 
of fact, 
 question of the form of this friction coefficient
and its dependence on the polymer length 
will be discussed elsewhere.

The property which will be studied here is 
the probability $P_{ch}(t)$ that the 
end-bead of the chain, which is not trapped at $t = 0$, 
remains not 
trapped until time $t$. Evidently, $P_{ch}(t)$ 
determines also  the probability 
that an initially unanchored chain remains 
completely mobile until time 
$t$. To calculate the time evolution of this property
we will proceed as follows: 
we will first present a formally 
exact expression for $P_{ch}(t)$
and show how its time dependence can be evaluated in the simplest 
case of a trivial chain
with $N = 0$, i.e. a single chemically active monomer. 
This will allow us 
to explain some basic methods and highlight 
the representative realizations of disorder
and  of the
monomer trajectories which support the intermediate 
and the long-time evolution
of $P_{ch}(t)$. Next, we will discuss
 the characteristic features of  dynamics of 
the polymer chain end-bead, which will
 allow then for a rather straightforward generalization
of the monomer trapping problem to a more complicated case with 
 the end-bead of a long polymer chain.

\section{A reminder on the monomer 
trapping problem in $d$-dimensional systems.}

To fix the ideas, it seems instructive to recall first the kinetic 
behavior of $P_{ch}(t)$ 
in the simplified case $N = 0$, i.e. the case of a chemically 
active monomer diffusing in $d$-dimensions and reacting with randomly 
placed, immobile perfect traps. It is intuitively clear that in our case 
with a chemically active monomer attached to a chain, we should retrieve at 
sufficiently large times (and up to some
renormalization of the diffusion coefficient)
  the behavior predicted for a single monomer, 
since 
for a finite chain random motion of any bead of the chain
ultimately converges to conventional diffusion 
with renormalized diffusion 
coefficient \cite{7}.  
At shorter times, however, substantial deviations 
should be observed 
because of essentially non-diffusive behavior of the 
end-bead, induced by internal relaxations of the chain.

The problem of kinetic description of chemical reactions between diffusive 
particles and immobile, randomly distributed traps has been widely 
discussed in the literature within the last two decades. Different 
analytical techniques have been elaborated, including an extension of the 
"optimal fluctuation" method \cite{bal}, different methods of bounds (see, 
e.g., \cite{don,pastur,gp,kh,burl}, Green functions approach \cite{burl}, 
field-theoretic treatments \cite{3}, as well as a variety of 
mean-field-type descriptions (see \cite{4,fix,deutch} and references 
therein). The interest to this problem was inspired by physical 
significance of the subject and, last but not least, by an early 
observation \cite{bal} that the long-time survival probability of diffusive 
particles exhibits highly non-trivial, fluctuation-induced behavior, which 
is relevant to the so-called Lifschitz singularities near the edge of the 
band in the density of states of a particle in quantum Lorentz gas.  Later 
works (see, e.g., \cite{burl,3}) have also pointed out relevance of the 
issue  to the problems of percolation,  self-avoiding random walks or 
self-attracting polymers, as well as anomalous behavior of 
the ground-state energy of the Witten's toy Hamiltonian 
of supersymmetric quantum 
mechanics \cite{sosiska}.  

Consider a single monomer, which diffuses (with diffusion coefficient 
$D_{0}$) in a $d$-dimensional volume $V$ with randomly placed $M$ traps, 
modeled as $d$-dimensional spheres of radius $\it a$. Positions of the 
traps are denoted by $\{{\bf R}_{j}\}$, where the subscript $j$ numerates the
traps, $j = 1, ... ,M$. 
The probability $\Psi_{mon}(t;\{{\bf R}_{j}\})$ 
that, for a given realization $\{{\bf R}_{j}\}$ of traps' 
distribution, a diffusive monomer will 
not encounter any of $M$  traps until time $t$ is given by (see, e.g. 
\cite{pastur,burl})
\be \label{probreal}  
\Psi_{mon}(t;\{{\bf R}_{j}\}) \; = \disp {\LARGE E}_{\Omega}\left \{ 
\exp\left [ - \int_{0}^{t} dt'  
\sum_{j = 1}^{M} W({\bf r}_{0}(t') - {\bf R}_{j}) \right ] \right\},
\ee
where the potential $W({\bf 
r}_{0}(t) - {\bf R}_{j})$ is  
the step-function, centered around the position of the $j$-th 
trap, such that
\be \label{potentialw}
W({\bf r}) \; = \; \left\{\begin{array}{ll}
\infty & \mathrm{|{\bf r}| \; \leq \; \it{a},}\\
0 & \mathrm{|{\bf r}| \; > \; \it{a},}
\end{array}
\right.
\ee
while
the symbol ${\Large E}_{\Omega}\{ \ldots \}$ 
denotes expectation 
on the complete set $\Omega$ of 
trajectories $\{{\bf r}_{0}(t)\}$ of a diffusive monomer.
We note parenthetically that Eq.(\ref{probreal}), which determines the 
monomer survival probability for a fixed configuration of traps, 
is not, of course,
only limited to the case when $\{{\bf r}_{0}(t)\}$ describe conventional
diffusion; Eq.(\ref{probreal}) is formally exact for $\em any$  
type of random or regular
motion, including the motion 
of the end-bead of a polymer chain, provided that 
the operator 
${\Large E}_{\Omega}$ is properly defined.

In what follows we will be  interested, however, not in the behavior of the
realization-dependent probability $\Psi_{mon}(t;\{{\bf R}_{j}\})$, but 
rather of its realization-averaged value $P_{mon}(t)$, defined as
\be \label{survival}
P_{mon}(t) \; = \; \Big\langle \Psi_{mon}(t;\{{\bf R}_{j}\}) 
\Big\rangle_{\{{\bf R}_{j}\}},
\ee
where the angle brackets denote here and henceforth the
averaging with respect to the distribution 
of traps positions 
$\{{\bf R}_{j}\}$. 
For random uncorrelated (Poisson) distribution of the traps such an 
averaging can be carried out 
straightforwardly \cite{pastur,burl}. Turning to 
the limit $V,M \to \infty$ and keeping
the ratio $M / V = n_{tr}$ fixed, one finds
\be \label{averaging}
\begin{array}{c}
\disp P_{mon}(t)  =  \disp  {\Large E}_{\Omega} \left\{
\prod_{j = 1}^{M} \left < \exp \left [ - \; \int_{0}^{t} dt' \;
W({\bf r}_{0}(t') - {\bf R}_{j}) \right ] \right> _{{\bf R}_{j}} 
\right \}   = \medskip \\ =  
\disp {\Large E}_{\Omega} \left\{
\; \exp \left [ - \; n_{tr} \; 
\int_{R^d} d{\bf R} \; \left ( 1 \; - \; \exp \left [ - \; \int_{0}^{t} dt' \;
W({\bf r}_{0}(t') - {\bf R}) \right ] \right ) \right ] \right \},
\end{array}
\ee
where the integral with the subscript $R^d$ in the last line of 
Eq.(\ref{averaging}) signifies that the integration 
extends over the entire $d$-dimensional space. It may be worth-while to note
that the function 
\be
1  -  \exp \left [ - \; \int_{0}^{t} dt' \;
W({\bf r}_{0}(t') - {\bf R}) \right ] = \; \left\{\begin{array}{ll}
1, & \mathrm{|{\bf r}_{0}(t') - {\bf R}| \; \leq \; \it{a}, \; t' \; \in \; [0;t],}\\
0, & \mathrm{|{\bf r}_{0}(t') - {\bf R}| \; > \; \it{a}, \; t' \; \in \; [0;t],}
\end{array}
\right.
\ee
is just the indicator function of the so-called Wiener sausage (see \cite{don,kac} 
for more
details) of size $a$, since it equals zero everywhere 
 except for the $a$-vicinity of any point of the particle
 trajectory ${\bf r}_{0}(t)$.
Consequently, the integral
\be
{\cal V}_{ws}\Big[{\bf r}_{0}(t)\Big] \; = \; \int_{R^d} d{\bf R} \; \left ( 1 \; - \; \exp \left [ - \; \int_{0}^{t} dt' \;
W({\bf r}_{0}(t') - {\bf R}) \right ] \right )
\ee
measures the volume swept out by diffusive spherical particle
 of radius $a$ during time $t$ 
for a particular
realization ${\bf r}_{0}(t)$ of its trajectory. 
In this regard, the realization-averaged
probability $P_{mon}(t)$ can be thought off as 
the moment generating function of the
volume of the Wiener sausage 
(see \cite{ber} for more details).  We also hasten to remark that
Eq.(\ref{averaging}) is quite general and can be also applied 
to describe the time evolution of
the
probability that 
the end-bead of the chain
does not encounter any 
of the traps until time $t$; to do this,
we have to define
the operator  ${\Large E}_{\Omega}$ as an expectation on the set of the end-bead
trajectories, whose properties, in general, will be different of those of a single monomer.

Expression
in the last line of Eq.(\ref{averaging}) 
determines an exact solution of the monomer 
trapping problem, which is valid at all times. Calculation
of $P_{mon}(t)$
amounts
now to performing averaging over 
the monomer trajectories. 
Such an averaging
procedure
has been extensively 
discussed in \cite{bal,don,pastur,burl},
 using different types of analytical approaches.
Here
 we intentionally choose
a method of bounds among other theoretical considerations,
 because
it gives us  a possibility not only to display in the most simple fashion 
the 
evolution of $P_{mon}(t)$ 
at intermediate and large times, but also
to highlight the representative 
monomer trajectories,
which support the corresponding 
decay pattern. This method   allows also for 
a rather straightforward computation of the
analogous probability not to encounter any of the traps until time $t$ 
in a 
more complicated situation with an active particle
attached to a polymer chain.

\subsection{Intermediate-time behavior of $P_{mon}(t)$.} 

Following 
Ref.\cite{burl}, a lower bound on $P_{mon}(t)$ in 
Eq.(\ref{averaging}) which  describes properly the
intermediate-time decay pattern can be readily 
found by making use of the 
Jensen-type inequality;   
this states that the averaged value of an
exponential of some 
random function $f$ is greater or equal to
the exponential of the averaged value of this function; i.e.
\be \label{jensen}
{\LARGE E} \Big\{ \exp\Big( - f\Big) \Big\} \; \geq \; 
\exp\Big( - {\LARGE E} \Big\{f\Big\}\Big) 
\ee
Hence, by setting 
\be
f = \disp \exp \left[ - \; n_{tr} \; 
\int_{R^d} d{\bf R} \; \left( 1 \; - \; \exp \left [ - \; \int_{0}^{t} dt' \;
W({\bf r}_{0}(t') - {\bf R}) \right ] \right ) \right ],
\ee
and applying the inequality in Eq.(\ref{jensen}), one finds that $P_{mon}(t)$  
can be bounded from 
below by 
\be \label{ineq}
\disp P_{mon}(t) \geq
\disp \exp \Big[ - \; n_{tr} \; {\LARGE E}_{\Omega}\Big\{{\cal V}_{ws}\Big[{\bf r}_{0}(t)\Big]
\Big\} \Big] 
\ee 
The time evolution
of the function ${\Large E}_{\Omega}\{{\cal V}_{ws}[{\bf r}_{0}(t)]
\}$ has been discussed, in particular, 
 in Refs.\cite{burl} and \cite{ber}. It has been shown
that expectation of the Wiener sausage 
volume obeys
\be \label{const}
{\LARGE E}_{\Omega}\left\{{\cal V}_{ws}[{\bf r}_{0}(t)]
\right \} \; = \; S_{d} \; \int_{a}^{\infty} r^{d-1} dr \; {\cal L}_{\lambda}^{-1}
\left[\frac{G(r;\lambda)}{\lambda \; G(a;\lambda)}\right]
 = \;  
\int_{0}^{t} dt' \; K_{d}(t'), 
\ee
where $S_{d}$
is the surface area 
of a $d$-dimensional sphere of radius $a$, $S_{d} = 2 \pi^{d/2}/\Gamma(d/2)$,
$\Gamma(x)$ being the Gamma-function, ${\cal L}_{\lambda}^{-1}[ \ldots]$ denotes the inverse
Laplace-transform operator, 
$K_{d}(t)$ is equal to the 
diffusive current through the surface of 
a $d$-dimensional immobile, adsorbing
 sphere of radius $a$, (i.e. $K_{d}(t)$ is the $d$-dimensional 
analog of the so-called Smoluchowski constant \cite{smol,ovch}), while 
\be
G(r;\lambda) \; = \; \int^{\infty}_0 dt \; \exp( - \lambda t) \; G(r;t),
\ee
where (the propagator)  $G(r;t)$ is the probability 
of finding a diffusive particle at distance $r$ 
from the starting point at
time $t$. Explicitly, one has that in the limit $t 
\gg a^2/D_{0}$ the expectation of the Wiener sausage 
volume or the time-integral of the
Smoluchowski-type constant show the following asymptotical behavior,
\be \label{int}
{\LARGE E}_{\Omega}\left\{{\cal V}_{ws}[{\bf r}_{0}(t)]
\right \} \; = \; \int^{t}_{0} dt' \; K_{d}(t') \; \approx \; 
\left\{\begin{array}{lll}
4 \pi a D_{0} t, & \mathrm{d = 3}\\
4 \pi D_{0} t/ln(4 D_{0} t/a^2), & \mathrm{d = 2}\\
4 (D_{0} t/\pi)^{1/2}, & \mathrm{d = 1,}
\end{array}
\right.
\ee
which thus depends  
on the spatial dimension $d$. 
This implies that in systems
of different dimensionality the typical number of intersections of the 
Wiener sausage behaves quite differently.  
Discussion of this point in terms of compact and non-compact exploration of space by
random walk trajectories, 
as well as the relation between
the mean 
volume of the Wiener sausage and the Smoluchowski-type rate constant, has been
presented first in \cite{pgg}. We note also that Eq.(\ref{const}) allows for computation
of the mean Wiener sausage volume for the trajectories of the end-bead, 
provided that its propagator
$G_{ch}(r;t)$ is known (see section IV.A).

Consequently, the bound based on the Jensen-type
inequality, leads to the result 
\be \label{smol}
\disp P_{mon}(t) \geq \disp P_{Smol}(t) = 
\disp \exp \left [ - \; n_{tr} \; \int^{t}_{0} dt' \; K_{d}(t') 
\right], 
\ee
where the expression in right-hand-side of Eq.(\ref{smol}), 
as first noted in \cite{burl}, 
is tantamount to the solution of the monomer 
trapping problem in terms of 
the celebrated Smoluchowski approach \cite{smol,ovch}. 
It is well-known from numerical studies
of the monomer trapping problem (see, e.g. \cite{numer})
that 
an approximation $P_{mon}(t) \approx P_{Smol}(t)$
describes fairly well 
the intermediate time behavior of the survival probability
 and fails to describe the decay properly only at very large 
times, 
when certain fluctuation effects comes into play.
The crossover times from the intermediate-time to the 
fluctuation-induced behavior will
be discussed in the next subsection.

\subsection{Long-time behavior of $P_{mon}(t)$.} 

Note now that the just outlined  
derivation \cite{burl} of  
 the Smoluchowski-type  result 
in Eq.(\ref{smol})   
demonstrates 
that an approximation $P_{mon}(t) 
\approx P_{Smol}(t)$
is equivalent to a certain 
assumption concerning the 
representative class 
of the monomer's
trajectories ${\bf r}_{0}(t)$, which is 
embodied into the 
Jensen inequality. 
This can be most easily seen if we set $F = exp( - f)$
and rewrite formally the inequality in Eq.(\ref{jensen}) as
\be \label{logjensen}
{\LARGE E}\Big\{F\Big\} \; \geq \; \exp\Big({\LARGE E}\Big\{ln\Big(F\Big)\Big\}\Big)
\ee
One notices now that the Jensen inequality 
Eq.(\ref{jensen}) bounds the averaged value of 
the functional $F$ by an exponential of the 
averaged logarithm of this functional; since logarithm is a very slowly varying function,
it is
generally believed that behavior of the averaged logarithm of some functional 
 is supported by
typical realizations of disorder. Consequently, 
one may claim that the Smoluchowski-type decay law $P_{mon}(t) 
\approx P_{Smol}(t)$, which describes properly the intermediate-time behavior,
is supported by $\em typical$ realizations 
of random walk trajectories ${\bf r}_{0}(t)$, i.e. 
such that ${\bf r}_{0}(t) \sim t^{1/2}$. In what follows we will thus 
refer to the
behavior supported by typical realizations of random walk trajectories
as the $\em mean-field-type$ behavior.

On the other
hand, 
at completely random placement of traps 
their local density 
will  
deviate throughout the volume 
from the volume-average value $n_{tr}$; 
there will be spatial regions
in which the density of traps is higher 
than $n_{tr}$, as well as regions containing no traps at all.
 One may thus expect that at large $t$ only those monomers 
will survive, which 
appear initially in sufficiently large trap-free regions 
and do not leave these regions until time $t$. 
Such restricted, atypical trajectories ${\bf r}_{0}(t)$, 
which also belong to the set $\Omega$ and which
are not taken into account in the Smoluchowski solution, 
 will contribute additively 
to the value of the probability $P_{mon}(t)$. Consequently, 
one may expect that 
the overall  probability $P_{mon}(t)$
will be of the form (see \cite{burl} for more details)
\be \label{sum}
\disp P_{mon}(t) \approx \disp P_{Smol}(t) \; + \; P_{fl}(t),
\ee
where the first term determines 
the behavior stemming out of typical realizations of random walk trajectories, 
while the second one represents 
the contribution of constrained 
trajectories entirely remaining within the
trap-free regions. 

Let us discuss now in more detail the contribution to 
the overall survival probability
stemming out of the constrained, atypical trajectories. 
As in the previous subsection, we will determine their contribution
evaluating a lower bound on $P_{mon}(t)$; for this purpose 
we adapt  to the path-integral formulation of the problem 
the approach developed
originally for three-dimensional systems in Ref.\cite{kac} and, independently,
for arbitrary $d$ in  Ref.\cite{gp}. 

We start again 
with the expression for the  survival probability of a monomer
diffusing in presence of  
traps fixed at positions $\{{\bf R}_{j}\}$, which is given by Eq.(\ref{probreal}).
The basic idea for evaluating the lower bound 
on the rhs of Eq.(\ref{probreal}) and subsequently, 
on $P_{mon}(t)$, is now as follows:
\begin{itemize}
\item Suppose that in calculating
an expectation of a positive definite functional
on a set of random walk trajectories we extend the 
integration not over the entire set of all possible trajectories
$\Omega$, but only over some subset $\omega$ of it, $\omega \in \Omega$. 
In doing so, we  evidently diminish the actual averaged value; consequently, 
one has 
\be \label{inequality}
\begin{array}{c}
\disp {\Large E}_{\Omega} \left \{ 
\prod_{j = 1}^{M} \exp \left [ - \; \int_{0}^{t} dt' \;
W({\bf r}_{0}(t') - {\bf R}_{j}) \right ] \right\} 
\; \geq \; \medskip \\
\geq  \disp \disp {\Large E}_{\omega} \left \{ 
\prod_{j = 1}^{M} \exp \left [ - \; \int_{0}^{t} dt' \;
W({\bf r}_{0}(t') - {\bf R}_{j}) \right ]  \right\}, 
\end{array}
\ee 
\item Let us define the subset $\omega$.
To do this, we first assume, without lack of generality, 
that a diffusive monomer is at the origin at $t = 0$ and that the 
nearest to the origin trap, say, the trap with $j = 1$, is at distance 
$R$. Now, we stipulate that 
$\omega$ is formed by such trajectories ${\bf r}_{0}(t)$
which start at the origin at $t = 0$ 
and during time interval $t$ do not cross the
surface of a 
$d$-dimensional sphere\footnote{Actually, it means that the number of
traps is effectively increased by introducing additional traps which cover
completely the
surface of a $d$-dimensional sphere of radius $R$ centered around
the origin. This certainly  can only diminish the bound.} of radius $R$. 
For the trajectories 
${\bf r}_{0}(t) \in \omega$, 
we evidently have that
\be \label{eq}
\disp \prod_{j = 1}^{M}  \exp \left [ - \; \int_{0}^{t} dt' \;
W({\bf r}_{0}(t') - {\bf R}_{j}) \right ] = 1,
\ee
since neither of such trajectories 
reaches any of the traps. Consequently, we find from Eq.(\ref{inequality}) that
\be \label{inequal}
{\Large E}_{\Omega} \left \{ 
\prod_{j = 1}^{M} \exp \left [ - \; \int_{0}^{t} dt' \;
W({\bf r}_{0}(t') - {\bf R}_{j}) \right ] \right\} 
\; \geq  \; {\cal P}(R; t) =  {\Large E}_{\omega}\left\{ 
 1  \right\}, 
\ee
in which ${\cal P}(R; t)$ denotes the measure of the
trajectories comprising the subset $\omega$. 
Eventually, one finds that the monomer survival probability is bounded by \cite{gp,kac}
\be \label{pmon}
\disp P_{mon}(t) \geq \disp {\cal P}(R; t) \; {\cal P}(R),
\ee
where ${\cal P}(R)$ is the probability of having a trap-free spherical 
void of radius $R$. 
\end{itemize}

For random uncorrelated (Poisson) distribution of traps the
 probability of finding a spherical 
cavity of radius $R$ completely devoid of traps is given by
\be \label{cavity}
{\cal P}(R) \sim \disp \exp( - n_{tr} {\it v}_{d} R^d),
\ee
where ${\it v}_{d} = 2 \pi^{d/2}/ d \; \Gamma(d/2)$ is the volume of
 a $d$-dimensional sphere of unit radius.
The measure  ${\cal P}(R; t)$ of trajectories comprising the subset $\omega$ 
equals the probability that a diffusive particle, which 
starts at the origin at $t = 0$,  does not hit the sphere at $|{\bf r}| = R$
 until time $t$. This probability  is given asymptotically by
\be \label{hit}
{\cal P}(R; t) \sim \disp \exp( - {\gamma}_{d} \frac{D_{0} t}{R^2}),
\ee
$\gamma_d$ being a dimensionless $d$-dependent number, 
$\gamma_{1,3} = \pi^2$ and $\gamma_2 \approx 
2.41$. Combining Eqs.(\ref{hit}),(\ref{cavity}) and (\ref{pmon}) one finds
\cite{gp,kac}:
\be \label{pmn} 
\disp P_{mon}(t) \geq \disp \exp\left(  - {\gamma}_{d} \frac{D_{0} t}{R^2} -  n_{tr} {\it v}_{d} R^d  \right),
\ee
which bound 
is valid for $any$ value of $R$. Hence, one has 
to choose such $R$ which provides the maximal value
to the rhs of Eq.(\ref{pmn}). The maximal lower bound obtains for 
\be \label{optimal}
R = R^{*}(t) = \disp \left( \frac{2 \gamma_d}{d {\it v}_d} 
\frac{D_{0} t}{n_{tr}}\right)^{1/(d + 2)}
\ee
Note now that $R^{*}(t)$ shows a 
slow 
growth with time. This implies that as time progresses 
larger and larger trap-free voids contribute most
importantly.  Eq.(\ref{optimal}) allows us to determine
 the representative
atypical realizations more
precisely: these are such realizations 
of random walk trajectories ${\bf r}_{0}(t)$
which grow with time not faster than
 $R^{*}(t) \sim t^{1/(d + 2)}$, i.e. are essentially
more spatially confined than the typical ones, for which 
one has ${\bf r}_{0}(t) \sim t^{1/2}$. That this should be the case is intuitively clear
since only those random walks survive at large times which do not make too large
excursions from their starting point.
 
The bound corresponding to $R = R^{*}(t)$ now
reads \cite{gp}:
\be \label{best}
\disp P_{mon}(t) \geq \disp P_{fl}(t) = \; 
\exp\left(  - \nu_d \; n_{tr}^{2/(d+2)} \; (D_{0} t)^{d/(d+2)} \right),
\ee
where $\nu_d$ are $d$-dependent numerical 
factors; in particular, for two- and three-dimensional systems 
$\nu_d$ is given respectively by 
$\nu_2 = 2 {\sl z}_{0} \sqrt{\pi}$, where 
${\sl z}_{0} \approx 2.405$ is the first zero of Bessel function $J_{0}(x)$,
and $\nu_3 = 5 \; 4^{1/5} \pi^{8/5}/3$
\cite{gp}. In what follows we will call the decay laws as in Eq.(\ref{best})
as $\em fluctuation-induced$, since such a behavior
consequences from the presence of fluctuations in the spatial distribution of traps 
and
respectively, 
from atypical realizations of random walk trajectories.

Gathering  Eqs.(\ref{smol}) and
(\ref{best}), we have now the following result for
the time evolution of $P_{mon}(t)$ in two-dimensional systems,
which will serve us in what follows as a point of reference,
\be \label{2dimmon}
P_{mon}(t) \; \approx \; \left\{\begin{array}{ll}
 \disp \exp\left( - 4 \; \pi \; n_{tr} \; D_{0} \; t/ ln(4 \; D_{0} \; t/a^2) \right), & \mathrm{{\it
a}^2/D_{0} < t <
t_{c}}\\
 \disp \exp\left( - \nu_2 \; (n_{tr} \; D_{0} \;  t)^{1/2}\right), 
& \mathrm{t > t_{c}}\\
\end{array}
\right.
\ee 
In Eqs.(\ref{2dimmon}) the time $t_{c}$ denotes the
crossover time, separating the Smoluchowski-type and the
fluctuation-induced kinetic regimes; this characteristic time is given by
\be \label{crossover}
t_{c} = \disp \frac{{\sl z}_{0}^2 \; a^2}{4 \; D_{0} \; \sigma_{tr}} 
\; ln^{2}\left(\frac{{\sl z}_{0}^2}{\sigma_{tr}}\right),
\ee
in which equation the parameter $\sigma = \pi \; a^2 \; n_{tr}$ determines the area of the surface
covered by traps.

Lastly, several comments on the magnitude of the 
crossover time and relative importance
of two regimes displayed in Eq.(\ref{2dimmon}) are in order.
On comparing
the terms in the exponentials of Eqs.(\ref{smol}) and 
(\ref{best}), we infer that 
atypical realizations
become progressively more important at such times when
$D_{0} t_{c}$ becomes greater than $\sigma_{tr}^{-2}$, 
$\sigma_{tr}^{-1} ln^2(\sigma_{tr})$ and $\sigma_{tr}^{-3/2}$ for 
one-, two- and three-dimensional systems, respectively.
Thus the crossover time may be quite large for systems 
in which the area covered by traps is low.
Moreover, there is another subtle circumstance
which makes
the fluctuation-induced tail, generally speaking, not pertinent for
real experimental systems.  Namely, the point is that
the amount of particles 
reacting at the intermediate-time Smoluchowski-type
kinetic stage is usually comparable to the total 
amount of particles in the system
such that up to the time $t_{c}$ only a few
particles are left. 
Numerical simulations of the monomer trapping kinetics (see \cite{numer}), 
which observe the
Smoluchowski-type regime and enter into 
the fluctuation-induced one, suggest that, in particular,
  in three-dimensions the fluctuation-induced regime
shows up when $P_{mon}(t)$ drops below $10^{-16}$,$10^{-25}$,$10^{-36}$ and
$10^{-80}$ for $\sigma_{tr}$ equal to $0.25$,$0.10$,$0.05$ and $0.01$, respectively.
In two-dimensions this should not be so dramatic as in 3D,
but still the value of
$P_{mon}(t_{c})$, which is defined as
\be \label{rdepth}
P_{mon}(t_{c}) \approx \disp \exp\left( - {\sl z}_{0}^2 \; ln\left({\sl
z}_{0}^2/\sigma_{tr}\right)\right),
\ee
can be very small due 
to the appearance of the factor $- ln(\sigma_{tr})$
in the exponent. Equation (\ref{rdepth}) suggests that in two-dimensions
the values of $P_{mon}(t)$ at the crossover will equal 
$10^{-8}$,$10^{-10}$,$10^{-12}$ and $10^{-16}$ for 
$\sigma_{tr}$ equal to $0.25$,$0.10$,$0.05$ and $0.01$, respectively.
We set out to show, however, that
for a chemically  active monomer  attached to a $long$
polymer chain the situation may change considerably such that
the value of $P_{mon}(t)$ at the crossover time $t_{c}$
will not be  that small.
 This resembles, in a way,  behavior predicted for the reverse counterpart
of the problem to be considered here - the trapping of diffusive monomers on traps
arranged in polymer chains. For this problem it has been also shown that the
fluctuation-induced behavior starts at much earlier times and most of particles
are trapped via the fluctuation-induced mechanism \cite{4}. We also note
that a similar effect of strong reinforcement of the magnitude of the
fluctuation-induced kinetics has been predicted for reactions involving active
particles attached to movable 
polymer chains in solution \cite{sergei}.

\section{Anchoring kinetics of a Rouse polymer
chain with a chemically active end-bead.}

In this section we will make use of the bounds, displayed
in Eqs.(\ref{smol}) and (\ref{best}), for computation of
the probability $P_{ch}(t)$ that an active 
monomer attached to one of the extremities of
a Rouse polymer chain 
still remains untrapped until time $t$. 
Consequently, as in the case with a single diffusive monomer,
we have to determine the form of
 two essential parameters:
the Smoluchowski-type constant 
describing reactions between the active end-bead
and traps, and the probability ${\cal P}(R; t)$ 
that the end-bead of a long polymer chain
remains within a trap-free cavity of radius $R$ until time $t$. 
While  behavior of $K_{d}(t)$ 
has been already discussed in the literature
within the context of reactions between particles attached to polymers  
(see Refs.\cite{4,pgg,sergei}), the 
form of the probability ${\cal P}(R; t)$ 
has not been considered so far. Clearly, computation of both $K_{d}(t)$ 
and ${\cal P}(R; t)$ requires the knowledge of the end-bead dynamics.

\subsection{Langevin dynamics of the end-bead of a Rouse polymer chain}

Let us briefly outline
the Langevin equation description
of the Rouse polymer chain dynamics 
 on a two-dimensional surface.  
In neglect of the excluded volume interactions and 
regarding the number $n$ of the bead in the chain as
continuous variable, $n \in [0,N]$, one has that 
dynamics of the vector of the n-th bead is governed by the
following Rouse-Langevin equation (see for more details \cite{7}):
\be \label{lang}
\eta \frac{\partial {\bf r}_n}{\partial t} =
\frac{2 T}{b^2} \frac{\partial^2 {\bf r}_n}{\partial n^2} \; + \; {\bf \zeta}_{n}(t)
\ee
where ${\bf \zeta}_{n}(t)$ are random forces, whose properties are described by
Eq.(\ref{noise}). Solution of Eq.(\ref{lang}), which corresponds to the
free boundary conditions at the chain extremities \cite{7}, i.e.
\be
\disp \frac{\partial {\bf r}_{n=0}(t)}{\partial n}=0; \quad
\frac{\partial {\bf r}_{n=N}(t)}{\partial n}=0,
\ee
can be written down as the Fourrier series of the form
\be \label{fourrier}
{\bf r}_n(t) = \disp \sum_{p = - \infty}^{\infty} {\bf X}_{p} 
\; \cos\left(\frac{p \pi n}{N}\right),
\ee
where the two-dimensional vectors ${\bf X}_{p}$ are
the normal coordinates of Eq.(\ref{lang}) (see \cite{7}). 
For further analysis it
suffices to know only their 
time correlation functions:
\be \label{normal}
\overline{X_{p,\alpha}(t) X_{q,\alpha'}(0)} \; = \; \delta_{p,q} \; 
\delta_{\alpha,\alpha'} \; \frac{D_{0} \tau_{R}}{N p^2} 
\; \exp\left( - \frac{p^2 t}{\tau_{R}} \right),
\ee
for $p > 0$, and 
\be \label{normalo}
\overline{(X_{0,\alpha}(t) - X_{0,\alpha}(0)) (X_{0,\alpha'}(t) 
- X_{0,\alpha'}(0))} \; = \;
\delta_{\alpha,\alpha'} \; \frac{2 D_{0} t}{N} 
\ee
for $p = 0$. In Eqs.(\ref{normal}) and (\ref{normalo})  
the symbols $\alpha$, $\alpha' = x,y$ denote, as before, 
 the Cartesian components of the
 normal coordinates and $\tau_{R}$ is the largest relaxation
time of the chain,  $\tau_{R} = b^2 N^2/2 \pi^2 D_{0}$. Physically, $\tau_{R}$ 
can be interpreted as being the time needed for some local defect,
e.g., kink, to spread out diffusively with diffusion constant $D_{0}$ 
along the arclength $b N$ of the chain.

Now, the property of interest for us
  is  the moment generating 
function for displacements
of the chain's end-bead, 
 which determines its propagator $G_{ch}(r;t)$
and thus contains the information we need
for calculation of $K_{d}(t)$ and ${\cal P}(R; t)$. 
The moment generation function is defined
as
\be \label{moment}
\Phi({\bf k}) = \disp \overline{\exp\Big(i \; {\bf k} \; {\bf r}_{n = 0}(t)\Big)}
\ee
The averaging in Eq.(\ref{moment}) over Gaussian white noise
can be performed straightforwardly, 
using Eq.(\ref{fourrier}) 
and the expressions for the time 
correlation functions of the normal coordinates.  One finds after some simple calculations

\be \label{momen}
\Phi({\bf k}) = \disp \exp\left( - \frac{{\bf k}^2}{4} \; 
\overline{{\bf r}_{n = 0}^{2}(t)} \right) = \;
\exp( - \; {\bf k}^2  D_{0} \tau),
\ee
where the
 effective "time" $\tau$ is a single-valued complicated
function of real time $t$:
\be \label{time}
\tau = \tau(t) = \disp \frac{t}{N} \; + \; \frac{2 \tau_{R}}{N} \; \sum_{p = 1} p^{-2} \left[1 \; - \; \exp\left( -
\frac{p^2 t}{\tau_{R}}\right)\right],
\ee
Eq.(\ref{momen}) implies that the probability of finding the end-bead at distance $r$ from the
starting point at time $t$ is a standard Gaussian function
\be \label{gauss}
G_{ch}(r;t) \; = \; \frac{1}{4 \pi D_0 \tau(t)} \; \exp\left( - \frac{r^2}{4 D_0 \tau(t)}\right)
\ee
Consequently, dynamics of the end-bead of a polymer can be considered as that
of a single monomer evolving in time $\tau$.

Asymptotical dependence of $\tau$ on $t$ 
can be readily calculated from Eq.(\ref{time}), which gives 
\be \label{timet}
\tau(t) \; \approx \; \left\{\begin{array}{ll}
t/N, & \mathrm{t > \tau_{R}}\\
b \left(t/ \pi D_{0}\right)^{1/2}, & \mathrm{t < \tau_{R}}
\end{array}
\right.
\ee
Eq.(\ref{timet}) signifies that $\tau$ scales with time 
differently depending whether $t$ is less or greater than
 the fundamental relaxation time $\tau_{R}$. 
We note also that 
the end-bead mean-square displacement (MSD) displays different 
time-behavior for $t \ll \tau_{R}$ and $t \gg \tau_{R}$; for  $t \ll \tau_{R}$ the motion of the end-bead
is due mainly to the internal relaxation of the chain. At such time scales
\be \label{msd}
\overline{{\bf r}_{n = 0}^{2}(t)} \; \approx \; 4 b
\left(\frac{D_{0} t}{\pi} \right)^{1/2}, 
\ee
which differs from conventional diffusion 
law in that the trajectory of the end-bead is
 spatially more confined. In the opposite time limit, $t \gg \tau_{R}$, the chain diffuses as one entity
and the end-bead trajectories follow the motion of the chain's center of mass. In this regime
the end-bead MSD obeys 
\be \label{msdt}
\overline{{\bf r}_{n = 0}^{2}(t)} \; \approx \; 
\frac{4 D_{0} t}{N}, 
\ee
i.e. conventional diffusion law
 with reduced diffusion coefficient
 $D = D_{0}/N$.

We close this subsection with the following conclusion. 
Dynamics of the end-bead of a Rouse polymer
is strongly influenced by the 
presence of the polymeric "tail"; 
at times less than the fundamental time 
$\tau_{R}$ the end-bead trajectories are 
spatially more confined compared 
to conventional diffusion
and its MSD shows a sublinear 
growth with time, Eq.(\ref{msd}). 
At greater times, 
the MSD grows linearly with time 
but the diffusion coefficient
is a factor of $1/N$ less than that 
for a monomer, Eq.(\ref{msdt}). Nonetheless, 
in view of the form of
 Eqs.(\ref{momen}) and (\ref{gauss}), 
the dynamics of the 
chain's end-bead 
can be considered
as that of an individual 
diffusive
monomer, which evolves in effective "time" $\tau$.
We will use this observation in 
what follows to estimate the time evolution of 
$K_{d}(t)$ and ${\cal P}(R; t)$. 

\subsection{Time evolution of $P_{ch}(t)$}

Consider first the contribution to $P_{ch}(t)$ stemming 
out of typical realizations of the 
end-bead trajectories, which problem amounts  to 
calculation of the Smoluchowski-type
constant $K_{d}(t)$ for an active group attached to a polymer chain. 
This question has been
 first addressed in \cite{pgg} and \cite{sergei} 
within the context of chemical 
reactions involving particles attached to movable polymer chains
 in solution (see for general discussion \cite{4}).  
It has been shown
that the form of the time-dependence of $K_{d}(t)$ depends 
on the time of 
observation $t$:  
for times $t$ less
than the largest
 relaxation time 
of the chain 
the Smoluchowski-type constant for a chemically active monomer
attached to 
a polymer chain should follow \cite{pgg}
\be \label{pgga}
K_{d}(t) \; \sim \; \frac{\left(\overline{{\bf r}_{n = 0}^{2}(t)}\right)^{d/2}}{t}
\ee
For a Rouse chain 
on a two-dimensional substrate, it gives, in particular,
\be
K_{2}(t) \; \sim \; b \; \left(\frac{D_{0}}{t}\right)^{1/2}
\ee
On the other hand, 
within the opposite limit, i.e. for $t \gg \tau_{R}$, 
when conventional diffusive motion with renormalized diffusion coefficient is
restored, one has,  (up to
the replacement $D_{0}  \to D_{0}/N$), 
conventional behavior 
as in Eq.(\ref{int}).

Let us now 
compute the time evolution of the 
Smoluchowski constant making use of
Eq.(\ref{const}), which will allow us to determine also the prefactors. 
Substituting the propagator in Eq.(\ref{gauss}) into Eq.(\ref{const})
we have that for 
the end-bead of a Rouse polymer chain the expectation of the Wiener sausage volume
 obeys
\be \label{win}
{\LARGE E}\Big\{{\cal V}_{ws}\Big[{\bf r}_{n = 0}(t)\Big]\Big\}
 =  \; 4 \pi D_0 \; {\cal L}_{\lambda}^{-1} \Big[
\frac{\int^{\infty}_0 dt \exp\Big( - \; \lambda t \; - 
\; a^2/4 D_0 \tau(t)\Big)}{\lambda \; 
\int^{\infty}_0 dt \; \tau^{-1}(t) 
\; \exp\Big( - \; \lambda t \; - \; a^2/4 D_0 \tau(t) \Big)}
\Big]
\ee
Consider the asymptotical behavior of 
${\Large E}\{{\cal V}_{ws}[{\bf r}_{n = 0}(t)]\}$ in the limits $t < \tau_R$ and 
$t > \tau_R$. Supposing that $\tau_R$ is large and 
setting $\tau(t) = b (t/\pi D_0)^{1/2}$ we find that the integrals in the nominator and
the denominator in Eq.(\ref{win}) 
behave as $1/\lambda$ and $\pi (D_0/\lambda)^{1/2}/b$, 
respectively. One finds then
\be
{\LARGE E}\Big\{{\cal V}_{ws}\Big[{\bf r}_{n = 0}(t)\Big]\Big\}
 = \int^t_0 dt' \; K_{2}(t') \; \approx \; 8 \; b \; \Big(D_{0} \; t/\pi\Big)^{1/2}
\ee
which holds 
for the time interval ${\it a}^2/D_{0} < t < \tau_{R}$.
 Next, we have that in the limit
$t > \tau_R$ the effective time $\tau(t) \approx t/N$. For such a time dependence
both integrals in  Eq.(\ref{win}) can be performed explicitly. This yields, upon some
straightforward calculations, the following result
\be \label{constchain}
{\LARGE E}\Big\{{\cal V}_{ws}\Big[{\bf r}_{n = 0}(t)\Big]\Big\}
 = \int^t_0 dt' \; K_{2}(t') \; \approx \; 
4 \; \pi \;  D_{0} \; t/N \; ln\Big(4 \;  D_{0} \; t/ a^2 \;
N\Big), 
\ee
which is valid for times $t > \tau_R$. Consequently, we find
 the time evolution of $P_{ch}(t)$ due to $\em typical$
 trajectories of the end-bead is defined by 
\be \label{smolchain}
P_{Smol}(t) \; \approx \; \left\{\begin{array}{ll}
\exp\left( - 8 \; b \; n_{tr} \; (D_{0} \; t/\pi)^{1/2}\right), &
\mathrm{{\it a}^2/D_{0} < t < \tau_{R}}\\
\exp\Big( - 4 \; \pi \; n_{tr} \;  D_{0} \; t/N \; ln(4 \;  D_{0} \; t/ a^2 \;
N)\Big), & \mathrm{t > \tau_{R}}
\end{array}
\right.
\ee
where the first line in Eq.(\ref{smolchain}) corresponds to the subdiffusive
motion of the chemically active end-bead, while the second one describes
the Smoluchowski-type decay pattern in the regime when the trajectories
of the end-bead start to follow the motion of the chain's center of mass.

To consider now the contribution to the decay of the atypical realizations
of the end-bead trajectories,
constrained not to leave until time $t$ the fluctuation trap-free voids,
we suppose that the probability that the end-bead remains with 
a circular trap-free  void of radius $R$ until time $t$ equals the probability
that an individual diffusive particle remains within such a cavity until time
$\tau(t)$.
Such an assumption gives
\be
{\sl P}(R;t) \; \approx \; \exp( - \gamma_{2} \frac{D_0 \tau(t)}{R^2})
\ee
Hence,  the contribution to $P_{ch}(t)$
due to atypical realizations  of the polymer end-bead trajectories
can be found by maximizing (with respect to $R$) the following expression
\be \label{cc}
P_{fl}(t) \; \approx \; max_{R} \left[ \exp( - n_{tr} {\it v}_2 R^2 \; - 
\; \gamma_{2} \frac{D_0 \tau(t)}{R^2})\right],
\ee
where the first term  determines the probability of 
having a trap-free circle of radius
$R$, while the second one gives the probability 
that the end-bead remains within such a
circle until time $t$.

Maximizing the rhs of Eq.(\ref{cc}), one
 readily finds that the value of $R$ which
yields the maximal value of the lower bound  is given by
\be
R \; = \; R^{*}(t) \; = \; \left(\frac{2 \gamma_2 }{2 {\it v}_2} \frac{D_0
\tau(t)}{n_{tr}}\right)^{1/4} \; \sim \; \left\{\begin{array}{ll}
\left(D_0 t/n_{tr}^2\right)^{1/8}, & \mathrm{t < \tau_{R}}\\
\left(D_0 t/N n_{tr}\right)^{1/4}, & \mathrm{t > \tau_{R}}
\end{array}
\right.
\ee
which implies that the representative atypical realizations of the end-bead
trajectories behave differently  depending on the scale of observation; for times $t$ less than $\tau_{R}$
atypical trajectories
of ${\bf r}_{0}(t)$ grow not faster than $t^{1/8}$, while within
the opposite limit they are constrained by ${\bf r}_{0}(t) < R^{*}(t) \sim t^{1/4}$.
Consequently, the contribution due to atypical realizations can be written down
explicitly as
\be \label{flchain}
P_{fl}(t) \; \approx \; \left\{\begin{array}{ll}
\exp\left( - \nu_2 \; (b \; n_{tr})^{1/2} \; (D_{0} \; t/\pi)^{1/4}\right), & \mathrm{{\it a}^2/D_{0} < t < \tau_{R}}\\
\exp\left( - \nu_2 \; (n_{tr} \;  D_{0} \; t/N)^{1/2} \right), & \mathrm{t > \tau_{R}}
\end{array}
\right.
\ee

Now, to construct an actual decay pattern describing  the anchoring
 kinetics in case
of an active monomer
attached to a polymer chain, we have to compare four different decay laws displayed in
Eqs.(\ref{smolchain}) and (\ref{flchain}) and calculate the corresponding crossover times.
From Eq.(\ref{sum}), which states that the mean-field and the fluctuation-induced decay laws complement each other,
we infer 
that different possible sequences of
 kinetic regimes may be observed, 
depending
on the magnitude of the 
parameters $n_{tr}$ and $N$, or, more precisely, depending on the relation between
$n_{tr}$ and $R_{g}$, where $R_{g} = b N^{1/2}$ is the chain's gyration radius.

Let us start with the case of sufficiently short chains and  low 
trap concentration, which limit 
is described by
 two inequalities: $a \; b \; n_{tr} \ll 1$ and
$n_{tr} R_{g}^2 \ll 1$.  
In this case we predict
 for the time evolution of $P_{ch}(t)$ the following sequence of regimes
\be \label{low}
P_{ch}(t) \; \approx \; \left\{\begin{array}{lll}
\exp\left( - 8 \; b \; n_{tr} \; (D_{0} \; t/\pi)^{1/2}\right), &
\mathrm{{\it a}^2/D_{0} < t < \tau_{R}}\\
\exp\Big( - 4 \; \pi \; n_{tr} \;  D_{0} \; t/N \; ln(4 \;  D_{0} \; t/ a^2 \;
N)\Big), & \mathrm{\tau_{R} < t < t_{c},}\\
\exp\left( - \nu_2 \; (n_{tr} \;  D_{0} \; t/N)^{1/2} \right), & \mathrm{t > t_{c}}
\end{array}
\right.
\ee
where the crossover time 
from the Smoluchowski-type behavior to the long-time
fluctuation-induced tail is given by
\be \label{cross}
t_{c} = \disp \frac{{\sl z}_{0}^2 \; a^2 N}{4 \; D_{0} \; \sigma_{tr}} 
\; ln^{2}\left(\frac{{\sl z}_{0}^2}{\sigma_{tr}}\right),
\ee 
i.e., is greater by a factor of $N$ than the corresponding crossover time for
the monomer trapping problem. 

Compare now the relative importance of the kinetic 
regimes displayed in Eq.(\ref{low}).
First of all, we note that the Smoluchowski-type regime associated with the
subdiffusive
behavior of the end-bead (first line in Eq.(\ref{low})) is not representative: 
although it can persist
over a wide time range  if $\tau_{R}$ is large, 
the number of active groups trapped
via this decay law is low, since $P_{mon}(t = \tau_{R}) \approx \exp(- 8 (n_{tr}
R_{g}^2)/(2 \pi)^{1/2} ) \approx 1$. Consequently, here we encounter essentially
the same behavior as in the monomer trapping
problem, (see Eq.(\ref{2dimmon})); we note that 
even $P_{mon}(t = t_{c})$ appears to be
 the same as
that in Eq.(\ref{rdepth}). Hence, in this limit the only effect of the
polymer tail is that 
the crossover time between 
the Smoluchowski-type and the fluctuation-induced behavior
gets increased.

Next we turn to the opposite limit of long chains, such that $n_{tr} R_{g}^2 \gg 1$,
but still assume that the trap concentration is sufficiently small and
 $a \; b \; n_{tr} \ll 1$.
Here the comparison of the decay laws in Eqs.(\ref{smolchain}) and 
(\ref{flchain}) suggests that the overall decay pattern is a succession of three
different regimes:
\be \label{high}
P_{ch}(t) \; \approx \; \left\{\begin{array}{lll}
\exp\left( - 8 \; b \; n_{tr} \; (D_{0} \; t/\pi)^{1/2}\right), &
\mathrm{{\it a}^2/D_{0} < t < t_{c} = {\sl z}_0^4 \pi^2/4^4 D_0 (b n_{tr})^2}\\
\exp\left( - \nu_2 \; (b \; n_{tr})^{1/2} (D_{0} \; t/\pi)^{1/4}
\right), & \mathrm{t_{c} < t < \tau_{R},}\\
\exp\left( - \nu_2 \; (n_{tr} \;  D_{0} \; t/N)^{1/2} \right), & \mathrm{t > \tau_{R},}
\end{array}
\right.
\ee
i.e., the Smoluchowski-type decay law
associated with the subdiffusive motion of the end-bead,
fluctuation-induced regime corresponding to the same anomalous motion of the end-bead
and lastly, 
fluctuation-induced regime corresponding to conventional diffusive motion (with
renormalized diffusion coefficient)
of the end-bead. 

We note now that in the limit $n_{tr} R_{g}^2 \gg 1$ 
 the crossover time $t_c$ from
the Smoluchowski-type to the fluctuation-induced behavior is given  by
\be \label{r}
t_c \approx \disp \left(D_{0} \; b^2 \; \sigma_{tr}^2\right)^{-1},
\ee
i.e. is proportional to the second inverse power of $\sigma_{tr}$ and can be large
if $\sigma_{tr} \ll 1$ (but still 
less than $\tau_{R}$). Taking the value of $P_{ch}(t)$  at the crossover time $t_c$
we find that  $P_{ch}(t_c) \approx 
\exp(- \pi {\sl z}_0^2/2)$, i.e. is independent of both the concentration of traps and 
the chain length. Since here  $P_{ch}(t_c) \approx  10^{- 4.5}$,  
we infer that 
such a mean-field-type regime
will be rather representative;  however, the amount of active end-beads of polymers
trapped via this  law will be substantially less as that for individual monomers. 
Consequently, contrary to the previously considered situation with short chains, here the amount of 
polymers anchored via 
 the fluctuation-induced decay law in the second line in
Eq.(\ref{high}) will be much higher.  Lastly, we note that 
the value of $P_{ch}(t)$  at the end of this kinetic stage is
\be
P_{ch}(t = \tau_{R}) \approx \disp \exp( - (n_{tr} R_{g}^2)^{1/2}),
\ee
i.e., is very small in the limit under consideration. Hence,
the regime described by the third line in  Eq.(\ref{high}) will not be observed.

We close our analysis with some remarks concerning the excluded-volume effects, 
which are discarded
here but may be certainly very important, especially  for two-dimensional systems. 
We note that excluded-volume
interactions impose additional
constraints on the 
end-bead dynamics which will  result apparently
in a slower, compared to Eq.(40), growth of the end-bead MSD
at the intermediate times
 (less than characteristic relaxation
 time of the 2D chain with excluded-volume interactions), and a stronger, 
compared to Eq.(41), $N$-dependence of the
end-bead diffusion coefficient in the long-time limit. 
Consequently, one expects that the behavior
in Eqs.(47.a)
and (51.a), 
(and accordingly, the intermediate kinetic stages 
in Eqs.(52.a), (54.a) and (54.b)),  
which are 
associated with internal relaxations of the chain and anomalous end-bead dynamics,
 would be described by stretched-exponential dependences 
with different (compared to  $1/2$ and $1/4$) values of the 
dynamical exponents. 
On the other hand,
the long-time behavior  as in Eqs.(47.b) and (51.b), 
(as well as that in Eqs.(52.b), (52.c) and (54.c)), 
will remain essentially unchanged
except that $N$ will enter in a somewhat different power. 
Lastly, we would like to mention that
our qualitative conclusions concerning
different realizations of disorder which support the intermediate-time 
and long-time behaviors will hold even in the presence of 
the excluded-volume interactions; in other words, we expect that also in this case 
the intermediate-time behavior of $P_{ch}(t)$
will be supported by typical realizations of the end-bead
trajectories, while the long-time behavior 
will stem out from the interplay between the internal relaxations 
of the chain and fluctuations in the
spatial distribution of traps.

\section{Conclusions.}

To summarize, we have studied Rouse-Langevin dynamics of an isolated 
polymer chain bearing a chemically active 
functional group, attached to one of
the chain's extremities, on
 a $2D$ solid
substrate with immobile, randomly placed 
chemically active sites (traps). 
For a particular situation
when the end-bead of a chain consisting of $N$ segments 
can be  
irreversibly trapped
 by any of these sites, which results in
a complete anchoring of the whole chain, we calculate
the time evolution of the probability $P_{ch}(t)$
that the initially unanchored chain
remains mobile until time $t$. We show that in case of 
relatively short chains the time evolution of $P_{ch}(t)$ 
proceeds essentially in the same way as that
for the monomer trapping problem; at intermediate times $P_{ch}(t)$ 
follows a standard-form $2D$
 Smoluchowski-type decay law $ln P_{ch}(t) \sim - t/N ln(t)$, which crosses over
at very large times to the fluctuation-induced stretched-exponential dependence
 $ln P_{ch}(t) \sim - (t/N)^{1/2}$,
stemming out of fluctuations in the spatial distribution of traps. 
We find next that for long chains
the kinetic behavior is quite different; 
here two representative kinetic stages
are the intermediate-time decay of the form
$ln P_{ch}(t) \sim - t^{1/2}$, while the long-time stage is 
described by the dependence 
$ln P_{ch}(t) \sim - t^{1/4}$. The intermediate-time decay 
is the
Smoluchowski-type law associated with
subdiffusive motion of the end-bead, while the long-time
fluctuation-induced decay  stems out of the interplay between 
fluctuations in traps distribution and 
internal relaxations of the chain.
 
\vspace{0.3in}

\section{Acknowledgments}

The authors wish to thank S.F. Burlatsky, A. Comtet, A. Blumen
 and Y. Fyodorov for helpful
discussions. This research was supported in part 
by the French-German  PROCOPE collaborative 
program.


\newpage
\begin{thebibliography}{99}
\bibitem{ala} U. Albrecht, A. Otto and P. Leiderer,  
Phys. Rev. Lett. {\bf 
68}, 3192 (1992); T. Ala-Nissila, S. Herminghaus, T. Hjelt 
and P. Leiderer, 
Phys. Rev.  Lett. {\bf 76},  4003 (1996)

\bibitem{alb} S. Herminghaus, U. Sigel and P. Leiderer, in: 
Proceedings of 
the Moriond Meeting on Short and Long Chains at Interfaces, 
Villars-sur-Ollon, January 1995

\bibitem{joel} A.M. Cazabat, M.P. Valignat, S. Villette, 
J. De Coninck and F. Louche, Langmuir {\bf 13}, 
4754 (1997)

\bibitem{caza} S. Bardon, M. Cachile, A.M. Cazabat, 
 X. Fanton, M.P. Valignat 
and S. Villette, Faraday Discuss. {\bf 104}, 307 (1996)

\bibitem{cazb} S. Villette, PhD Thesis, Universit\'e Paris VI, 1996, 
unpublished

\bibitem{cazc} M.P. Valignat, G. Oshanin, S. Villette, A.M. Cazabat and 
M. Moreau, Phys. Rev. Lett. {\bf 80}, 5377 (1998) and cond-mat/9804270

\bibitem{cazd} A.M. Cazabat, J. De Coninck, S. Hoorelbeke, 
M.P. Valignat and S. Villette, 
Phys. Rev. E 
{\bf 49}, 4149 (1994)

\bibitem{tiberg} M.P. Valignat, N. Fraysse, A.M. Cazabat and F. Heslot, 
Langmuir {\bf 9}, 601
(1993)

\bibitem{yoon} B.G. Min, J.W. Choi, H.R. Brown, D.Y. Yoon, T.M. O'Connor and
M.S.Jhon, Tribology Letters {\bf 1}, 225
(1995)

\bibitem{nechaev} S.K. Nechaev, J. Mod. Phys. B {\bf 4}, 1809 (1990);
A. Grosberg and S.K. Nechaev, Adv. Polymer Sci. {\bf 106}, 1 (1993)

\bibitem{stauffer} G.M. Foo, R.B. Pandey and D. Stauffer, Phys. Rev. E {\bf 53},  3717  
(1996)

\bibitem{weiss} F. den Hollander and G.H. Weiss, 
in: Contemporary Problems in Statistical
Physics, ed. G.H. Weiss (SIAM, Philadelphia, 1994)

\bibitem{kehr} J.W. Haus and K.W. Kehr, Phys. Rep. {\bf 150}, 263 (1987); 
 K.W. Kehr and T. Wichmann, cond-mat/9602121

\bibitem{4} G. Oshanin, M. Moreau and S.F. Burlatsky, Adv. Colloid Inter. Sci. 
{\bf 49} (1994), 1

\bibitem{gomer} R. Gomer, Rep. Prog. Phys. {\bf 53}, 917 (1990)

\bibitem{7} M. Doi and S.F. Edwards, The Theory of Polymer Dynamics,
(Oxford Univ. Press, Oxford, 1986)

\bibitem{bal} B.Ya. Balagurov and V.T. Vaks, Sov. Phys. JETP {\bf 38}, 968 
(1974);

\bibitem{don} M.D. Donsker and S.R.S. Varadhan, Comm. Pure Appl. Math. {\bf 
28}, 525 (1975)

\bibitem{pastur} L.A. Pastur, Theor. Math. Phys. {\bf 32}, 88 (1977)

\bibitem{gp} P. Grassberger and I. Procaccia, J. Chem. Phys. {\bf 77}, 6281 
(1982)

\bibitem{kh} R.F. Kayser and J.B. Hubbard, Phys. Rev. Lett. {\bf 51}, 6281 
(1982)

\bibitem{burl} S.F. Burlatsky and A.A. Ovchinnikov, Sov. Phys. JETP {\bf 65}, 
908 (1987)

\bibitem{3} T.C. Lubensky, Phys. Rev. A {\bf30}, 2657 (1984); S.R. Renn, 
Nucl. Phys. B {\bf 275}, 273 (1986); Th.M. Nieuwenhuizen, Phys. Rev. Lett. 
{\bf 62}, 357 (1989)

\bibitem{fix} M. Fixman, J. Chem. Phys. {\bf 81}, 3666 (1984)

\bibitem{deutch} S.F. Burlatsky, O. Ivanov and J.M. Deutch, J. Chem. Phys. 
{\bf 97}, 156 (1992)



\bibitem{sosiska} C. Monthus, G. Oshanin, A. Comtet and S.F. Burlatsky, Phys. 
Rev. E {\bf 54}, 231 (1996)

\bibitem{kac} M. Kac and J.M. Luttinger, J. Math. Phys. {\bf 15}, 183 (1974)

\bibitem{ber} A.M. Berezhkovskii, Yu.A. Makhnovskii and R.A. Suris, 
J. Stat. Phys. {\bf 57}, 333 (1989); {\bf 65}, 1025 (1991)

\bibitem{smol} M. von Smoluchowski, Z. Phys. Chem. {\bf 92}, 
129 (1917)


\bibitem{ovch} D. Calef and J.M. Deutch, Annu. Rev. Phys. Chem. {\bf 34}, 493 (1983); 
G.H. Weiss, J. Stat. Phys. {\bf 42}, 3 (1986)

\bibitem{pgg} P.G. de Gennes, J. Chem. Phys. {\bf 76}, 3316 (1982)


\bibitem{numer} see, e.g., A. Blumen, J. Klafter and G. Zumofen,  
in: Optical Spectroscopy of Glasses, ed.: 
I. Zschokke, (Reidel Publ., Dordrecht, 1986); 
S. Havlin, M. Dishon, J.E. Kiefer and G.H. Weiss, Phys. Rev.
Lett. {\bf 53}, 407 (1984);  J. Anlauf, Phys. Rev.
Lett. {\bf 52}, 1845 (1984) and references therein

\bibitem{sergei} S.F. Burlatsky, Sov. Phys. Doklady
{\bf 286}, 155 (1986)

\end{thebibliography}
\end{document}